\documentclass[runningheads]{llncs}
\usepackage[T1]{fontenc}
\usepackage{graphicx}
\usepackage{lmodern}
\usepackage{amsfonts}
\usepackage{amsmath}
\usepackage{booktabs}
\usepackage{microtype}
\usepackage{makecell}
\usepackage{array}
\newcolumntype{C}{>{\centering\arraybackslash}p{2cm}}  
\usepackage{graphicx}
\usepackage[percent]{overpic}
\usepackage{caption} 
\usepackage{float}
\usepackage{xcolor}
\usepackage{amssymb}  
\usepackage{multirow} 
\usepackage[hidelinks, urlcolor=blue]{hyperref}

\begin{document}
%

\title
{SLaM-DiMM: Shared Latent Modeling for Diffusion Based Missing Modality Synthesis in MRI}

\titlerunning{SLaM-DiMM}

\author{
  Bhavesh Sandbhor$^{1}$, Bheeshm Sharma$^{2}$, Balamurugan Palaniappan$^{3}$
}

\institute{
  $^{1}$Department of MEMS, $^{2,3}$Department of IEOR, IIT Bombay, India. \\
  \email{\{$^{1}$22b2446, $^{2}$bheeshmsharma, $^{3}$balamurugan.palaniappan\}@iitb.ac.in}
}

\maketitle              
\begin{abstract}
Brain MRI scans are often found in four modalities, consisting of T1-weighted  with and without contrast enhancement (T1ce and T1w), T2-weighted imaging (T2w), and Flair.
Leveraging complementary information from these different modalities enables models to learn richer, more discriminative features for understanding brain anatomy, which could be used in downstream tasks such as anomaly detection.
However, in clinical practice, not all MRI modalities are always available due to various reasons. This makes missing modality generation a critical challenge in medical image analysis. In this paper, we propose SLaM-DiMM, a novel missing modality generation framework that harnesses the power of diffusion models to synthesize any of the four target MRI modalities from other available modalities. Our approach not only generates high-fidelity images but also ensures structural coherence across the depth of the volume through a dedicated coherence enhancement mechanism. Qualitative and quantitative evaluations on the BraTS-Lighthouse-2025 Challenge dataset demonstrate the effectiveness of the proposed approach in synthesizing anatomically plausible and structurally consistent results. Code is available at  \href{https://github.com/BheeshmSharma/SLaM-DiMM-MICCAI-BraTS-Challenge-2025}{\textcolor{blue}{\texttt{this link}}}.
\keywords{Brain MRI Synthesis  \and Diffusion Models \and Medical Imaging.}
\end{abstract}

\section{Introduction}
Missing modality generation \cite{baltruschat2024brasyn2023challengemissing} in brain MRI has emerged as a critical research area due to its impact on the training and performance of deep learning models used for anatomical localization and pixel-level annotation. Brain MRI data is typically acquired in multiple modalities, including T1-weighted imaging with and without contrast enhancement (T1ce and T1w), T2-weighted imaging (T2w), and Flair, each capturing distinct tissue properties and pathological features. While deep learning models have achieved strong performance in anomaly detection and analysis tasks, their effectiveness is often compromised in clinical settings where one or more modalities may be missing due to protocol differences or other reasons. The absence of modalities limits the model's access to complementary information crucial for accurate detection and characterization of anomalous regions. 

To address the challenge, we propose SLaM-DiMM: Shared Latent Modeling for Diffusion Based Missing Modality Synthesis in MRI. SLaM-DiMM is a novel framework designed to synthesize any target brain MRI modality from other available input modalities, enabling robust and flexible cross-modal image generation. Our architecture follows an encoder-decoder paradigm, in which a shared encoder maps the multi-channel input into a compact shared latent representation. This latent representation is then refined through a Latent Diffusion Model (LDM) \cite{rombach2022highresolutionimagesynthesislatent}-based bottleneck. The enhanced latent representation is subsequently decoded by modality-specific decoders to generate high-fidelity outputs in the target MRI modality.
Furthermore, we introduce a coherence enhancement network (CEn) that explicitly enforces 3D spatial consistency across depth, improving volumetric smoothness in the reconstructed 3D volumes. Synthesis results obtained using SLaM-DiMM on BraTS-Lighthouse 2025 Challenge dataset \cite{li2024braintumorsegmentationbrats}, highlight its superior ability to preserve fine textures and tissue contrast across diverse imaging modalities.

\section{Related Work}

Dar et al. \cite{dar2019image} proposed pGAN and cGAN architectures to synthesize missing modalities using adversarial learning for multi-contrast MRI translation, and Dalmaz et al. introduced ResViT \cite{dalmaz2022resvit}, a transformer-based model that integrates convolutional and vision transformer blocks for realistic synthesis. In 3D wavelet diffusion model (WDM), a diffusion model is directly applied on wavelet-decomposed images to preserve high-frequency details while enabling full-resolution image generation \cite{friedrich2024wdm}. Further conditional Wavelet Diffusion Model (cWDM) extends the WDM architecture by conditioning on the available modalities to enable effective cross-modality MRI synthesis \cite{friedrich2024cwdm}. Recently, HF-GAN \cite{cho2024unified} was used as a baseline to synthesize 2D MR images, incorporating a hybrid fusion encoder, a channel attention-based feature fusion module, a modality infuser, and a CNN decoder; additionally, a 3D U-Net-based refiner was used to enhance the quality of the synthesized slices by leveraging volumetric context \cite{cho2024two}. SwinUNETR \cite{hatamizadeh2022swinunetrswintransformers}, a hybrid architecture combining Swin Transformers and CNNs, was also applied to MRI synthesis and showed strong performance \cite{pang2025multi}. UNETR, where convolutional encoders are replaced with transformers to model long-range spatial dependencies in 3D segmentation, has achieved state-of-the-art results on multiple benchmarks, and enhances the quality of 3D MRI by capturing global contextual information during synthesis \cite{cho2024two}.

Recently, diffusion models \cite{ho2020denoising} have gained popularity for medical image generation, with successful applications in segmentation and anomaly detection \cite{wyatt2022anoddpm}.  Denoising Diffusion Probabilistic Models (DDPMs) \cite{ho2020denoising} have demonstrated strong performance in image synthesis tasks; however, they are computationally expensive due to their operation in high-dimensional pixel space. To address this, Rombach et al. \cite{rombach2022highresolutionimagesynthesislatent} introduced Latent Diffusion Models (LDMs), which operate in a learned latent space to reduce training cost while preserving high image quality. LDMs can also be used for paired image-to-image translation, such as MRI sequence conversion. These prior works form the basis for our proposed method, which integrates 2D intensity-based generation and 3D volumetric enhancements, minimizes slice-based artifacts and improves 3D consistency, resulting in enhanced segmentation performance.

\section{SLaM-DiMM Methodology}
This section contains the discussion of our proposed model SLaM-DiMM which includes missing modality generation (MMG) and coherence enhancement (CEn). Our approach is inspired by recent advances in deep generative modeling, particularly Latent Diffusion Models (LDMs) \cite{rombach2022highresolutionimagesynthesislatent} and Variational Autoencoders (VAEs) \cite{Kingma_2019}.

\subsection{Missing Modality Generation (MMG)}

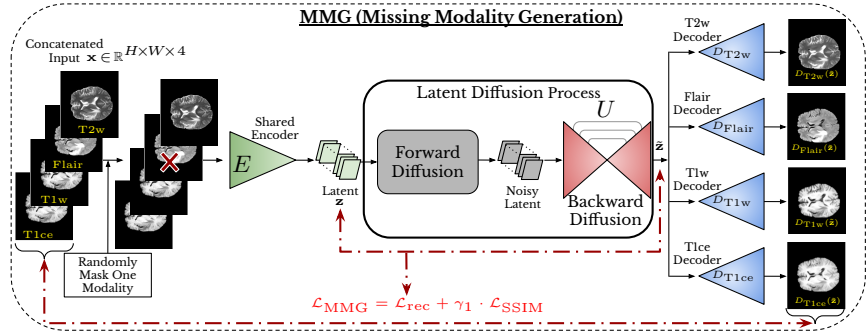
\begin{figure}[h!]
	\centering
	\begin{overpic}[width=0.95\linewidth,grid=false]{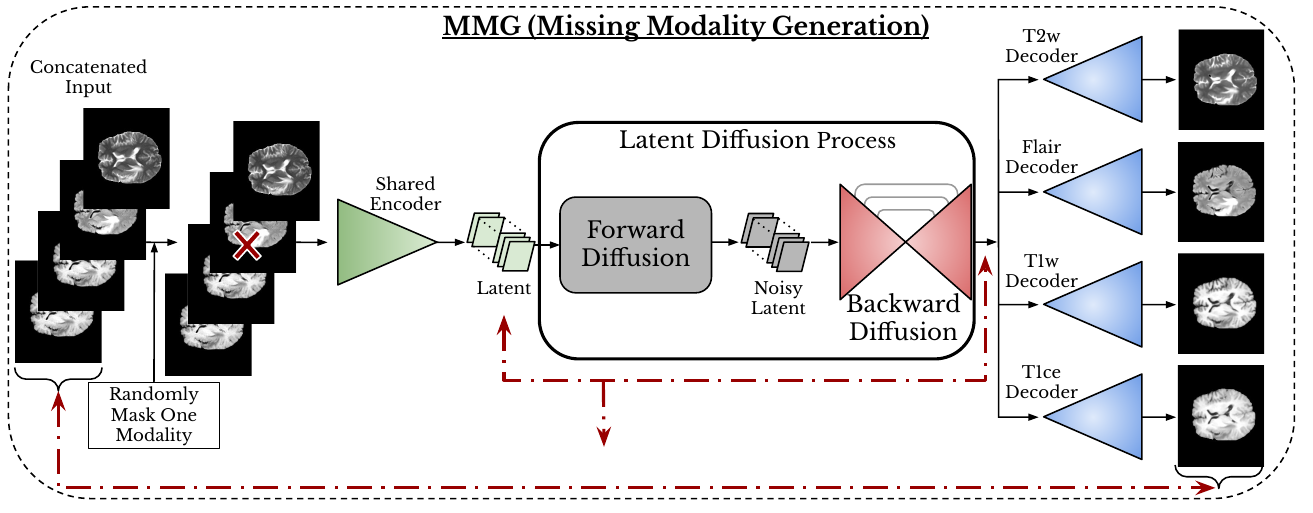}
		\put(9.3,32){\tiny{\( \mathbf{x}\!\in\!\mathbb{R}^{H\!\times\!W\!\times 4} \)}}
		\put(8.3,23.7){\scalebox{0.8}{\textcolor{yellow}{\tiny{\(\text{T2w}\)}}}}
		\put(5.4,19.7){\scalebox{0.8}{\textcolor{yellow}{\tiny{\(\text{Flair}\)}}}}
		\put(4,15.7){\scalebox{0.8}{\textcolor{yellow}{\tiny{\(\text{T1w}\)}}}}
		\put(2,11.7){\scalebox{0.8}{\textcolor{yellow}{\tiny{\(\text{T1ce}\)}}}}
		
		\put(26,19.2){\small{\( E\)}}
		
		\put(81,32.7){\scalebox{0.8}{\tiny{\(D_\text{T2w}\)}}}
		\put(81,24.2){\scalebox{0.8}{\tiny{\(D_\text{Flair}\)}}}
		\put(81,15.7){\scalebox{0.8}{\tiny{\(D_\text{T1w}\)}}}
		\put(81,7.1){\scalebox{0.8}{\tiny{\(D_\text{T1ce}\)}}}
		
		\put(67.8,25.5){\small{\( U\)}}
		
		\put(37.8,15.5){\tiny\(\mathbf{z}\)}
		\put(74.5,21.7){\tiny\(\tilde{\mathbf{z}}\)}

		\put(90.5,30.2){\scalebox{0.6}{\tiny\textcolor{yellow}{\(D_\text{T2w}(\tilde{\mathbf{z}})\)}}}
		\put(90.1,21.7){\scalebox{0.6}{\tiny\textcolor{yellow}{\(D_\text{Flair}(\tilde{\mathbf{z}})\)}}}
		\put(90.5,13.2){\scalebox{0.6}{\tiny\textcolor{yellow}{\(D_\text{T1w}(\tilde{\mathbf{z}})\)}}}
		\put(90.5,4.2){\scalebox{0.6}{\tiny\textcolor{yellow}{\(D_\text{T1ce}(\tilde{\mathbf{z}})\)}}}
		
		\put(35,3.7){\tiny\textcolor{red}{\(\mathcal{L}_{\text{MMG}} = \mathcal{L}_{\text{rec}} + \gamma_1 \cdot \mathcal{L}_{\text{SSIM}}\)}}
	\end{overpic}
	\caption{SLaM-DiMM architecture for multi-modality synthesis}
	\label{fig:Any-to-any}
\end{figure}

Our MMG architecture consists of a shared encoder, a Latent Diffusion Model (LDM)-inspired bottleneck, and four modality-specific decoders, as illustrated in Figure~\ref{fig:Any-to-any}. This design enables the model to learn a unified latent representation that captures high-level semantic features common across different MRI modalities while still allowing for modality-specific reconstruction through dedicated decoding. By learning a shared, modality-invariant latent space, the architecture facilitates flexible and accurate cross-modal synthesis, enabling the generation of any target modality from other available input modalities.
\vspace{-0.5cm}
A 3D brain MRI volume is represented as $ X_{m} \in \mathbb{R}^{H \times W \times D} $, where $ H $, $ W $, $ D $ denote the height, width, depth of volume, respectively. Each volume $X_m$  is associated with a specific imaging modality $ m \in \{\text{T1w}, \text{T1ce}, \text{Flair}, \text{T2w}\} $.

For computational efficiency during training, we operate on 2D axial slices extracted from the 3D volumes. Specifically, we process individual slices $ X_{m}^{f} \in \mathbb{R}^{H \times W} $ independently, where $ f \in \{1, 2, ..., D\} $ denotes the slice index within a volume. This 2D-slice based approach significantly reduces memory consumption and training time while still preserving the diagnostic quality of the data. During inference, the model progressively processes all slices, generating the corresponding reconstructed 2D slice outputs, which are then concatenated along the depth dimension to reconstruct the full 3D MRI volume.

\textbf{Encoding Stage:}
The encoder serves as the first stage of the pipeline, which extracts a lower-dimensional latent representation from the input 2D slices. It takes as input a 4-channel tensor \( \mathbf{x} \in \mathbb{R}^{H \times W \times 4} \), where each channel corresponds to one of the four MRI modalities. During training, we simulate missing modalities by randomly masking one of the four input channels with zeros (see Figure \ref{fig:Any-to-any}). This forces the model to learn robust, shared representations that can infer the missing modality based on the context provided by the remaining three.

The encoder maps the input to a lower-dimensional latent representation
$\mathbf{z} = E(\mathbf{x}) \in \mathbb{R}^{h \times w \times d}$
where \( E(\cdot) \) denotes the encoder function, and \( \mathbf{z} \) is a latent feature map of reduced spatial dimensions \( h \times w \) and channel depth \( d \). The encoder follows the design of the downsampling path of the OpenAI UNet \cite{nichol2021improveddenoisingdiffusionprobabilistic}.

\textbf{Latent Diffusion Bottleneck Stage:} 
Once the latent \( \mathbf{z} \) is obtained, it is passed through a Latent Diffusion Model (LDM) \cite{rombach2022highresolutionimagesynthesislatent}-inspired diffusion-based bottleneck. This bottleneck plays a critical role in refining the latent representation by introducing a stochastic transformation that enhances the semantic richness and generalization capability of the shared latent space across all imaging modalities. 

The forward diffusion process, gradually corrupts the latent features \( \tilde{\mathbf{z}}_0 = \mathbf{z} \) with Gaussian noise, producing \( \tilde{\mathbf{z}}_t \) over a series of time steps \( t = 1, \dots, T, \) using:
$
q(\tilde{\mathbf{z}}_t | \tilde{\mathbf{z}}_{t-1}) = \mathcal{N}\left(\tilde{\mathbf{z}}_t; \sqrt{1 - \beta_t} \tilde{\mathbf{z}}_{t-1}, \beta_t \mathbf{I} \right)
$, 
where \( \beta_t \) controls the variance of the noise at timestep \(t\).
The full forward process from \( \tilde{\mathbf{z}}_0 \) to \( \tilde{\mathbf{z}}_t \) can then be expressed in closed form as
$
q(\tilde{\mathbf{z}}_t | \tilde{\mathbf{z}}_0) = \mathcal{N}\left(\tilde{\mathbf{z}}_t; \sqrt{\bar{\alpha}_t} \tilde{\mathbf{z}}_0,\ (1 - \bar{\alpha}_t)\mathbf{I} \right)
$, 
where \( \alpha_t = 1 - \beta_t \) and \( \bar{\alpha}_t = \prod_{s=1}^t \alpha_s \). This allows us to directly sample \( \tilde{\mathbf{z}}_t \) from \( \tilde{\mathbf{z}}_0 \) without computing intermediate steps. After \( T \) steps, \( \tilde{\mathbf{z}}_T \) becomes approximately pure Gaussian noise \( \mathcal{N}(0, \mathbf{I}) \). 

The reverse diffusion process aims to reconstruct the original latent features by learning to iteratively denoise the corrupted representation. It is defined as
$
p_\theta(\tilde{\mathbf{z}}_{t-1} | \tilde{\mathbf{z}}_t) = \mathcal{N}\left(\tilde{\mathbf{z}}_{t-1}; \mu_\theta(\tilde{\mathbf{z}}_t, t), \Sigma_\theta(\tilde{\mathbf{z}}_t, t)\right)
$, 
where the mean \( \mu_\theta \) and covariance \( \Sigma_\theta \) are parameterized by a neural network (OpenAI UNet \cite{nichol2021improveddenoisingdiffusionprobabilistic}),  with parameters \( \theta \). The network is conditioned on the time step \( t \), enabling it to learn distinct denoising behaviors at different stages of the process.
This process results in refined, shared latent space features \( \tilde{\mathbf{z}} \), which capture modality-invariant features useful for reconstructing all four MRI modalities.

\textbf{Modality-specific Decoding Stage:}
To reconstruct different modalities from the shared latent representation \( \tilde{\mathbf{z}} \), we employ four independent decoders corresponding to the four modalities T1w, T1ce, Flair, and T2w. These decoders allow the model to preserve modality-specific characteristics while still benefiting from a shared modality-invariant latent representation. For each modality \( m \in \{\text{T1w}, \text{T1ce}, \text{Flair}, \text{T2w}\} \), the corresponding decoder \( D_{m}(\cdot) \) maps the shared latent representation back to the image domain:
$
\widehat{\mathbf{x}}_{m} = D_{m}(\tilde{\mathbf{z}}) \in \mathbb{R}^{H \times W}
$
where \( \widehat{\mathbf{x}}_{m} \) is the reconstructed output for modality \( m \). The decoders follow the upsampling path of the OpenAI UNet architecture, without skip connections, to ensure that the model relies solely on the shared latent features for reconstruction.

\textbf{Training and Inference:}
To ensure accurate reconstruction of all four MRI modalities while preserving the integrity of the latent representation, we define a composite reconstruction loss that includes both image-space and latent-space components as:
$$
\mathcal{L}_{\text{rec}} = \underbrace{\sum_{m=1}^4 \left\| (\widehat{\mathbf{x}}_{m} - \mathbf{x}_{m}) \cdot (1 + \lambda_1 \cdot \mathbf{S})\right\|_2^2}_{\text{Spatially weighted Image-space reconstruction loss}} + \underbrace{\lambda_2 \cdot \left\| \tilde{\mathbf{z}} - \mathbf{z} \right\|_2^2}_{\text{Latent-space consistency loss}}
$$
where \(\widehat{\mathbf{x}}_{m} = D_{m}(U(E(\mathbf{x})))\), \( E(\cdot) \) denotes the shared encoder, \( U(\cdot) \) represents the latent diffusion bottleneck, \( D_{m}(\cdot) \) denotes the modality-specific decoder for MRI modality \( m \in \{\text{T1w}, \text{T1ce}, \text{Flair}, \text{T2w}\} \), and $ \left\| \cdot \right\|_2^2 $ denotes the element-wise squared L2 loss. 

The first term defines the spatially weighted image-space reconstruction loss. $ \mathbf{S} \in \{0,1\}^{H \times W} $ is a binary segmentation mask given in the dataset as the ground truth, which highlights regions of interest (anomalous regions). This mask is used to spatially weigh the reconstruction loss, placing higher emphasis on anomalous regions where accurate reconstruction is most vital. \( \lambda_1 \) is a hyperparameter that controls the degree of importance placed on these anomalous regions. Note that the ground truth mask $\mathbf{S}$ is only used in training; during inference, ground truth mask is not necessary. 

The second term introduces a latent-space consistency loss to regularize the diffusion bottleneck and prevent excessive information loss during the stochastic refinement process.
In the second term, $ \mathbf{z} = E(\mathbf{x})$ is the initial latent representation obtained from the encoder, $ \tilde{\mathbf{z}} $ is the refined latent representation after passing through the diffusion bottleneck, $\lambda_2$ controls the strength of the constraint on latent information preservation. This term ensures that the diffusion process enhances, rather than distorts, the meaningful content of the latent space.

To further improve the perceptual quality of the generated images, we incorporate an SSIM-based loss. This loss measures the structural similarity between the generated outputs and the corresponding ground truth slices:
$
\mathcal{L}_{\text{SSIM}} = \sum_{m=1}^4 \left( 1 - \text{SSIM}\left(\widehat{\mathbf{x}}_{m}, \mathbf{x}_{m}\right) \right)$.
This term helps preserve anatomical structures of the synthesized images, especially in complex regions such as anomalies or tissue boundaries.

The total training objective is a weighted combination of the above two losses:
$
\mathcal{L}_{\text{MMG}} = \mathcal{L}_{\text{rec}} + \gamma_1 \cdot \mathcal{L}_{\text{SSIM}},
$
where $ \gamma_1 $ is a hyperparameter that balances the contributions of the SSIM loss component. This composite loss ensures that the model learns to reconstruct missing modalities accurately while maintaining high structural fidelity.

\subsection{Coherence Enhancement (CEn)}
\begin{figure}[ht]
    \centering
    \begin{overpic}[width=0.9\textwidth]{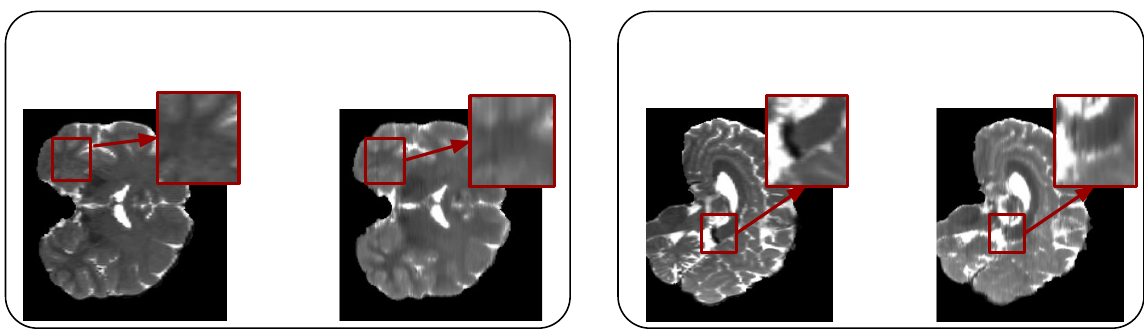}
        \put(3,20){\small $\text{Original}$}
        \put(33,20){\small $\text{MMG}$}
        \put(56,20){\small $\text{Original}$}
        \put(84,20){\small $\text{MMG}$}
        \put(18,24){\large $\text{Coronal}$}
        \put(72,24){\large $\text{Sagittal}$}
    \end{overpic}
    \caption{An example of MMG-generated slices from the coronal plane and sagittal plane views. A comparison between the original and MMG-generated samples reveals inter-slice inconsistencies, evident as vertical blurry lines in the MMG-generated slices, which become more pronounced upon zooming out.}
    \label{fig:CE_example}
\end{figure}
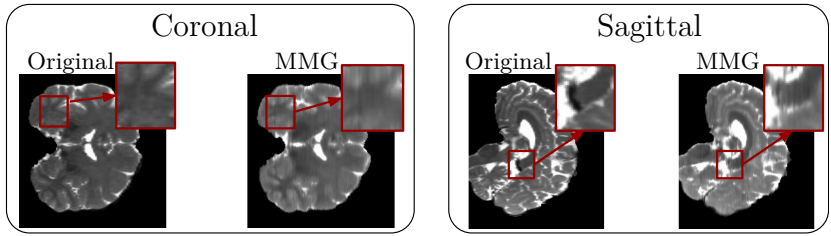

Following the generation of 2D MRI slices across all modalities and their subsequent concatenation into a 3D volume, we observed a degradation in volumetric coherence. Figure \ref{fig:CE_example} shows an example of an original Glioma slice and an MMG-generated slice from the coronal and sagittal views, where the lack of coherence is observable. This is primarily due to the frame-level training paradigm, which does not explicitly model spatial dependencies between adjacent slices. As a result, the reconstructed volumes may exhibit inter-slice intensity mismatches, structural discontinuities, and loss of anatomical consistency along the depth axis.

\vspace{-0.5cm}
\vspace{-0.5cm}
To address this limitation, 
motivated by \cite{cho2024two}, we introduce a Coherence Enhancement Network (CEn) based on the 3D-UNETR \cite{hatamizadeh2021unetrtransformers3dmedical}, which integrates transformer-based self-attention mechanisms with a 3D-UNet backbone, enabling it to capture long-range contextual dependencies in 3D space. This enables the network to effectively reduce artifacts and improve anatomical consistency throughout the reconstructed volume.

In our CEn training setup, we utilize the initial reconstructed volume $ \widehat{\mathbf{V}}_m \in \mathbb{R}^{H \times W \times D}$, from MMG  for each modality \(m\), where $ H, W, D $ are the spatial dimensions. The network processes the volume in a fully volumetric manner, refining the inter-slice transitions and aligning anatomical structures across the depth dimension. 
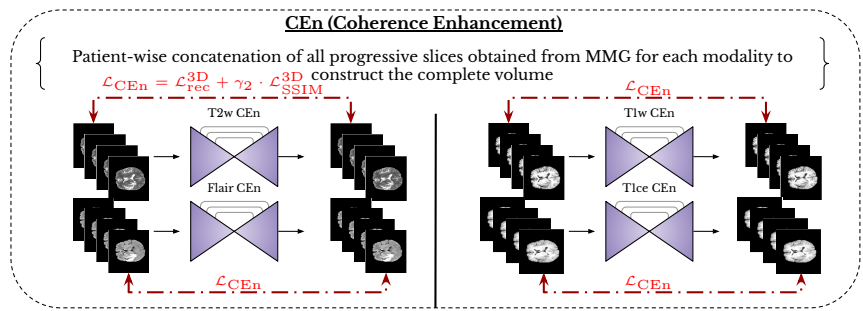
\begin{figure}[t!]
    \centering
    \begin{overpic}[width=0.95\linewidth,grid=false]{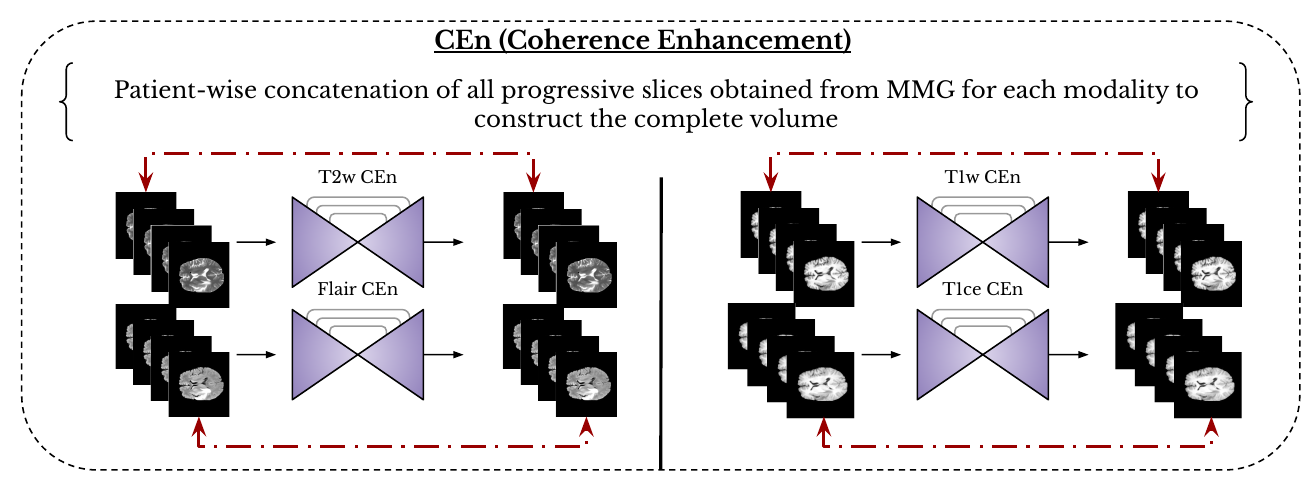}
        \put(12,26.5){\tiny\textcolor{red}{\(\mathcal{L}_{\text{CEn}} = \mathcal{L}^{\text{3D}}_{\text{rec}} + \gamma_2 \cdot \mathcal{L}^{\text{3D}}_{\text{SSIM}}\)}}
        \put(25,3.7){\tiny\textcolor{red}{\(\mathcal{L}_{\text{CEn}}\)}}
        \put(72,3.7){\tiny\textcolor{red}{\(\mathcal{L}_{\text{CEn}}\)}}
        \put(72,26){\tiny\textcolor{red}{\(\mathcal{L}_{\text{CEn}}\)}}
    \end{overpic}
    \caption{CEn architecture designed to enhance inter-slice consistency in reconstructed 3D MRI volumes.}
    \label{fig:CEn}
\end{figure}
To manage computational complexity, the synthesized 3D volume of size $ H \times W \times D $ is divided into overlapping sub-volumes of size $ H \times W \times \frac{D}{s} $, where \(s\) is a positive integer referred to as the subvolume factor. This factor controls the depth of each sub-volume. A sliding-window approach with a stride \(\frac{D}{2s}\) along the depth dimension is employed to ensure sufficient overlap between adjacent sub-volumes. 

The CEn model is trained using a composite reconstruction loss that combines both pixel-wise similarity and perceptual similarity (captured using SSIM), between the 3D sub-volumes. The overall CEn loss is defined as a weighted combination:
$
\mathcal{L}_{\text{CEn}} = \mathcal{L}^{\text{3D}}_{\text{rec}} + \gamma_2 \cdot \mathcal{L}^{\text{3D}}_{\text{SSIM}}
$
where \( \gamma_2 \in [0,1] \) balances the contribution of the SSIM loss component.

\section{Experimental Details and Results}

\subsection{Experimental Setup}
All experiments were conducted on an NVIDIA RTX A6000 GPU with 48 GB of memory, using the PyTorch framework. In the MMG model, we employ a configuration of 3 blocks for encoders and decoders in $E$, $D_m$, and $U$, with each block comprising 5 residual sub-blocks. For the diffusion process, time steps are uniformly sampled from the interval \([1, T]\) during training, where \(T = 1000\). During inference, we used DDPM sampling with a fixed time step \(t_{\text{test}} = \frac{T}{2} = 500\). A linear noise schedule is applied for \(\beta_t\), ranging from \(1 \times 10^{-4}\) to \(2 \times 10^{-2}\). The MMG model is trained for a maximum of 1600 epochs, where each epoch consists of processing 500 randomly selected image slices. Optimization is performed using the Adam optimizer with a learning rate of \(10^{-4}\) and a batch size of 4.  The reconstruction loss combines multiple components, weighed by hyperparameters $\lambda_1 = 4$ and $\lambda_2 = 2$.
To select the optimal value for hyperparameter $\gamma_1$, we evaluated the MMG model performance using the SSIM metric with values $\gamma_1 \in \{0.1, 0.25, 0.5, 0.75\}$. The corresponding SSIM scores were 94.59, 94.66, 94.78, and 94.51, respectively. The highest SSIM was achieved at $\gamma_1 = 0.5$, which was therefore used in all experiments.

For the CEn network, a subvolume factor $s = 10$ is used to extract 16-slice subvolumes from each 3D volume, with a stride of $\frac{D}{2s} = 8$ along the depth axis. Training is performed for up to 250 epochs with a batch size of 2, and each epoch includes 100 randomly sampled 3D volumes. Similarly, we conducted experiments to tune $\gamma_2$ over the same set of values: $\{0.1, 0.25, 0.5, 0.75\}$. The obtained SSIM scores were 94.84, 94.18, 94.44, and 94.24, respectively. Based on these results, $\gamma_2 = 0.1$ yielded the best performance and was selected for the final model.

\subsection{Dataset Details and Preprocessing}
The BraSyn-2025 dataset \cite{li2024braintumorsegmentationbrats,baid2021rsnaasnrmiccaibrats2021benchmark,menze2014multimodal,bakas2017advancing,bakas2017segmentation,spyridon2017segmentation,moawad2024braintumorsegmentationbratsmets} is derived from a combination of three publicly available datasets: BraTS-GLI 2023, BraTS-METS 2023, and BraTS-MENINGIOMA. It represents a retrospective multi-institutional collection of brain tumor multi-parametric MRI (mpMRI) scans acquired under standard clinical conditions. These scans vary significantly in image quality and characteristics due to differences in scanner hardware, imaging protocols, and institutional practices, thereby introducing realistic clinical heterogeneity. Expert neuroradiologists have manually reviewed and validated the ground truth annotations for all tumor sub-regions, ensuring high-quality segmentation masks.

The training dataset consists of 1,251 glioma patient volumes and 238 metastasis patient volumes. For validation, the dataset includes 219 glioma and 31 metastasis cases. Each volume contains four MRI modalities: T1-weighted (T1w), T2-weighted (T2w), Flair, and contrast-enhanced T1-weighted  (T1ce). Training dataset further contains ground-truth segmentation masks. During training, complete sets of all four modalities are provided in BraSyn-2025 data. In contrast, for validation and testing, one of the modalities is randomly omitted in each sample to evaluate the model's ability to synthesize the missing modality.

Prior to training, we apply intensity normalization by clipping values to the range between the 0.5th and 99.5th percentiles to mitigate outliers. The top and bottom 15 axial slices are removed, and the remaining 3D volumes are decomposed into 2D axial slices, which are saved as individual images for training.

\subsection{Quantitative Results}
The quantitative performance of the proposed MMG (Missing Modality Generation) and MMG$+$CEn (MMG with Coherence Enhancement) models is evaluated on the BraSyn2025 validation dataset, which includes 219 Glioma and 31 Metastasis patient volumes. Each model is tasked with synthesizing one of the four MRI modalities (T1w, T2w, Flair, T1ce). The performance metric used is the Structural Similarity Index Measure (SSIM), computed across the entire 3D volume for each modality synthesis task.

\vspace{-0.5cm}
\definecolor{darkgreen}{rgb}{0.0, 0.5, 0.0}
\begin{table}[ht]
    \centering
    \caption{SSIM (\%) comparison of MMG and MMG$+$CEn for missing modalities. The symbol $\textcolor{red}{\small\textbf{\texttimes}}$ indicates missing modality, and $\textcolor{darkgreen}{\small\textbf{\checkmark}}$ represents available modality.}
    \label{tab:combined_ssim}
    \begin{tabular}{@{\hspace{1cm}}c@{\hspace{0.2cm}}c@{\hspace{0.2cm}}c@{\hspace{0.2cm}}c@{\hspace{1cm}}c@{\hspace{0.2cm}}c@{\hspace{0.2cm}}c@{\hspace{0.2cm}}c}
        \toprule
        \multicolumn{4}{c}{\textbf{Modality}} &
        \multicolumn{2}{c}{\textbf{Glioma}} & 
        \multicolumn{2}{c}{\textbf{Metastasis}} \\
        \cmidrule(lr){1-4} \cmidrule(lr){5-6} \cmidrule(lr){7-8}
        T1w & T2w & Flair & T1ce & MMG & MMG$+$CEn & MMG & MMG$+$CEn \\
        \midrule
        \textcolor{red}{\small\textbf{\texttimes}} & \textcolor{darkgreen}{\small\textbf{\checkmark}} & \textcolor{darkgreen}{\small\textbf{\checkmark}} & \textcolor{darkgreen}{\small\textbf{\checkmark}} & 94.96 &  94.81  & 91.61 &  91.97  \\
        \textcolor{darkgreen}{\small\textbf{\checkmark}} & \textcolor{red}{\small\textbf{\texttimes}} & \textcolor{darkgreen}{\small\textbf{\checkmark}} & \textcolor{darkgreen}{\small\textbf{\checkmark}} & 93.82 &  93.86  & 89.61 & 90.39  \\
        \textcolor{darkgreen}{\small\textbf{\checkmark}} & \textcolor{darkgreen}{\small\textbf{\checkmark}} & \textcolor{red}{\small\textbf{\texttimes}} & \textcolor{darkgreen}{\small\textbf{\checkmark}} & 91.75 & 91.94 & 89.41 & 90.29 \\
        \textcolor{darkgreen}{\small\textbf{\checkmark}} & \textcolor{darkgreen}{\small\textbf{\checkmark}} & \textcolor{darkgreen}{\small\textbf{\checkmark}} & \textcolor{red}{\small\textbf{\texttimes}} & 92.44 & 92.26 &  89.67 & 89.98  \\
        \bottomrule
    \end{tabular}
\end{table}

\begin{table}[ht]
	\centering
	\caption{Dice and HD95 scores for MMG and MMG+CEn methods on Glioma and Metastasis datasets across labels.}
	\label{tab:gli_met_scores}
	\begin{tabular}{l@{\hspace{0.2cm}}c@{\hspace{0.2cm}}c@{\hspace{0.2cm}}c@{\hspace{0.2cm}}c@{\hspace{0.2cm}}c@{\hspace{0.2cm}}c@{\hspace{0.2cm}}c@{\hspace{0.2cm}}c}
		\hline
		& \multicolumn{4}{c}{Glioma} & \multicolumn{4}{c}{Metastasis} \\
		\cmidrule(lr){2-5} \cmidrule(lr){6-9}
		& \multicolumn{2}{c}{MMG} & \multicolumn{2}{c}{MMG+CEn} & \multicolumn{2}{c}{MMG} & \multicolumn{2}{c}{MMG+CEn} \\
		\cmidrule(lr){2-3} \cmidrule(lr){4-5} \cmidrule(lr){6-7} \cmidrule(lr){8-9}
		& Dice$\uparrow$ & HD95$\downarrow$ & Dice$\uparrow$ & HD95$\downarrow$ & Dice$\uparrow$ & HD95$\downarrow$ & Dice$\uparrow$ & HD95$\downarrow$ \\
		\hline
		Tumor Core & 0.7601 & 3.6651 & 0.7301 & 4.0226 & 0.5571 & 11.3043 & 0.5046 & 11.9173 \\
		Whole Tumor & 0.8993 & 2.1149 & 0.8876 & 2.4600 & 0.8211 & 10.7675 & 0.8155 & 5.1284 \\
		  Enhancing Tumor& 0.7873 & 3.3372 & 0.7702 & 3.6991 & 0.7912 & 13.6090 & 0.7836 & 7.9839 \\
		\hline
	\end{tabular}
\end{table}
\vspace{-0.5cm}
The results for the MMG and MMG$+$CEn models are summarized in Table~\ref{tab:combined_ssim}, where for the MMG model, the highest SSIM scores are observed for T1w synthesis in Glioma cases (94.96\%), and the lowest for Flair in Metastasis cases (89.41\%). On average, both MMG and MMG+CEn perform slightly better on Glioma samples compared to Metastasis across all modalities, likely due to the larger number of Glioma samples in the training set. The MMG+CEn model further refines the output of MMG model by enforcing spatial coherence and structural fidelity of the synthesized volume across the depth axis. The results show a marginal improvement in SSIM scores for Metastasis cases compared to the MMG model.

Further, Table \ref{tab:gli_met_scores} presents the Dice and HD95 distance scores for individual segmentation labels (Tumor Core (TC), Whole Tumor (WT), Enhancing Tumor (ET)) on the validation dataset. The segmentation inference is performed using pseudo-labels produced by the BraTS orchestrator model \cite{kofler2025bratsorchestratordemocratizing}, on generated data where, for each sample, one modality is randomly masked and reconstructed using both the MMG and MMG+CEn models. 
The results show that for Label 3 (ET), both MMG and MMG+CEn achieve comparable performance across the Glioma and Metastasis datasets. For Label 2 (WT), there is a slight performance dip in the Metastasis dataset compared to Glioma. However, for Label 1 (TC), there is a more significant drop in performance between the two datasets, suggesting that generating or segmenting tumor core is more challenging in metastasis cases.


\subsection{Qualitative Results}
\begin{figure}[!t]
	\centering
	\begin{overpic}[width=\textwidth]{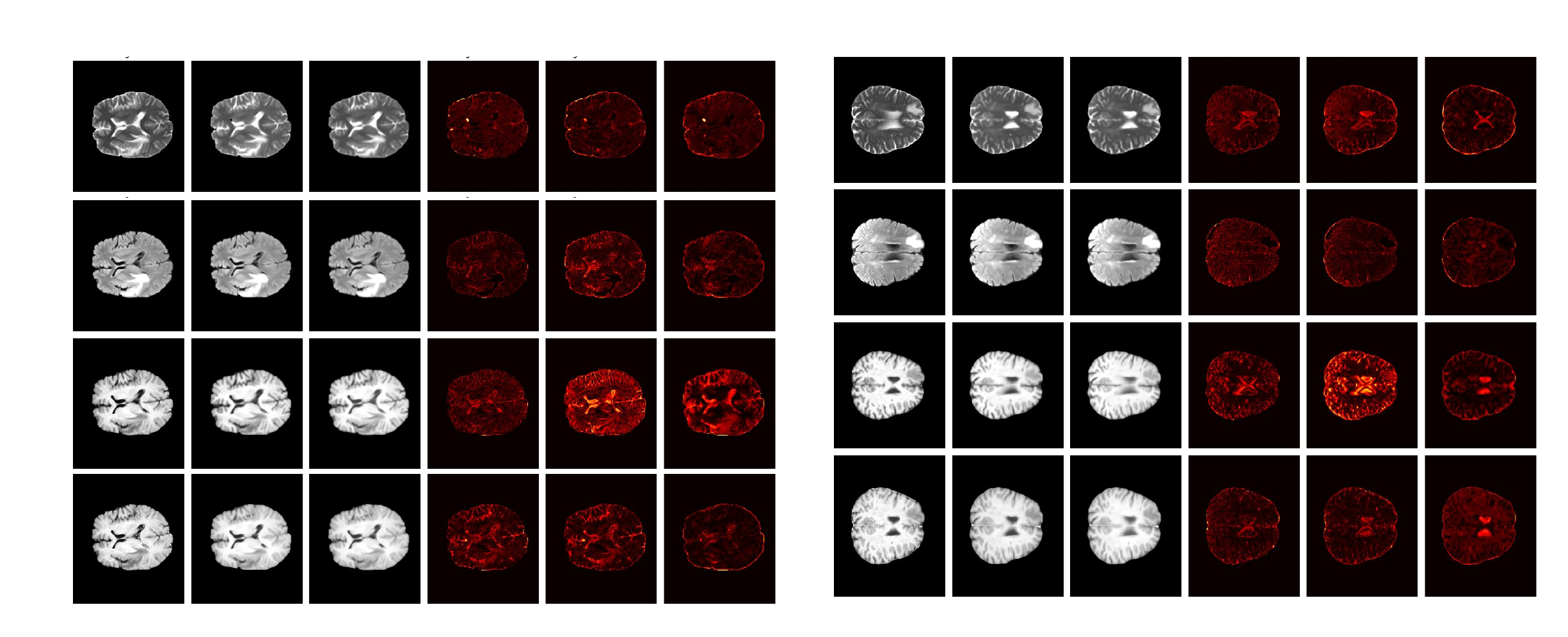}
		\put(12,40){\small $\text{Glioma Patient Samples}$}
		\put(60,40){\small $\text{Metastasis Patient Samples}$}
		\put(5,37.5){\tiny $\text{Original}$}
		\put(14,37.5){\tiny $\text{MMG}$}
		\put(22,37.5){\tiny $\text{CEn}$}
		\put(30,37.5){\tiny $\text{O-M}$}
		\put(38,37.5){\tiny $\text{O-C}$}
		\put(46,37.5){\tiny $\text{M-C}$}
		\put(53,37.5){\tiny $\text{Original}$}
		\put(62,37.5){\tiny $\text{MMG}$}
		\put(70,37.5){\tiny $\text{CEn}$}
		\put(78,37.5){\tiny $\text{O-M}$}
		\put(86,37.5){\tiny $\text{O-C}$}
		\put(94,37.5){\tiny $\text{M-C}$}
		\put(0,32){\tiny $\text{T2w}$}
		\put(0,23){\tiny $\text{Flair}$}
		\put(0,14){\tiny $\text{T1w}$}
		\put(0,5){\tiny $\text{T1ce}$}
	\end{overpic}
	\caption{Qualitative comparison between MMG and MMG$+$CEn. Columns "O–M", "O–C", and "M–C" denote difference maps between the original and MMG-generated slices, the original and MMG$+$CEn-generated slices, and the MMG and MMG$+$CEn generated slices, respectively. These maps highlight the structural modifications introduced by each model.}
	\label{fig:qualitative}
\end{figure}
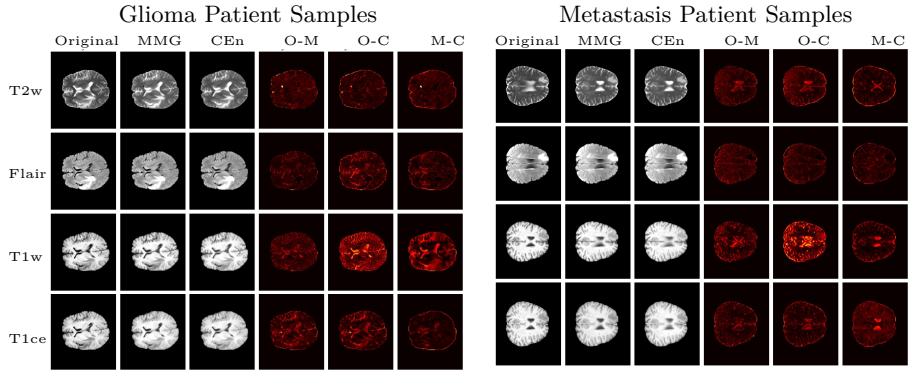

Figure~\ref{fig:qualitative} presents a qualitative comparison of the synthesized MRI modalities for patients from glioma (left) and metastasis (right) abnormalities. 
To evaluate generation quality and structural fidelity, difference maps are provided as "O–M", "O–C" and "M-C" in Figure \ref{fig:qualitative}, visualizing pixel-wise deviations between the synthesized results and the ground truth. The "O–M" map (Original vs. MMG) reveals only minor discrepancies across all modalities, indicating that MMG achieves high anatomical consistency and preserves fine structural details with minimal distortion. In contrast, the "O–C" map (Original vs. MMG+CEn) exhibits larger intensity differences in the T1w modality, suggesting that the CEn module induces more noticeable changes relative to the original anatomical structure. The "M–C" map (MMG vs. MMG+CEn) further illustrates these model-level differences. While reconstructions for T2w, Flair, and T1ce remain largely consistent between the two variants, T1w shows higher residual, reflecting significant modifications introduced by the coherence enhancement component.

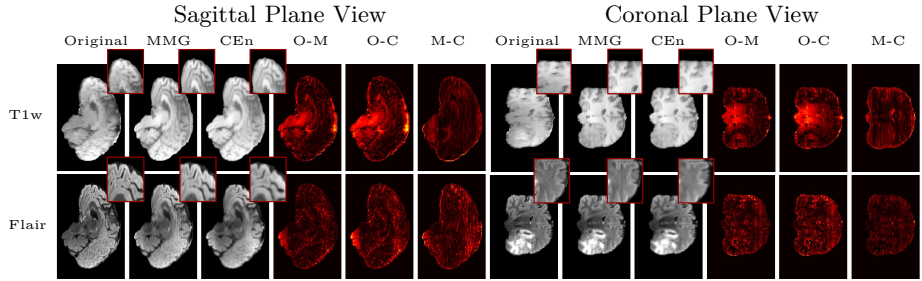
\begin{figure}[!t]
	\centering
	\begin{overpic}[width=1\textwidth]{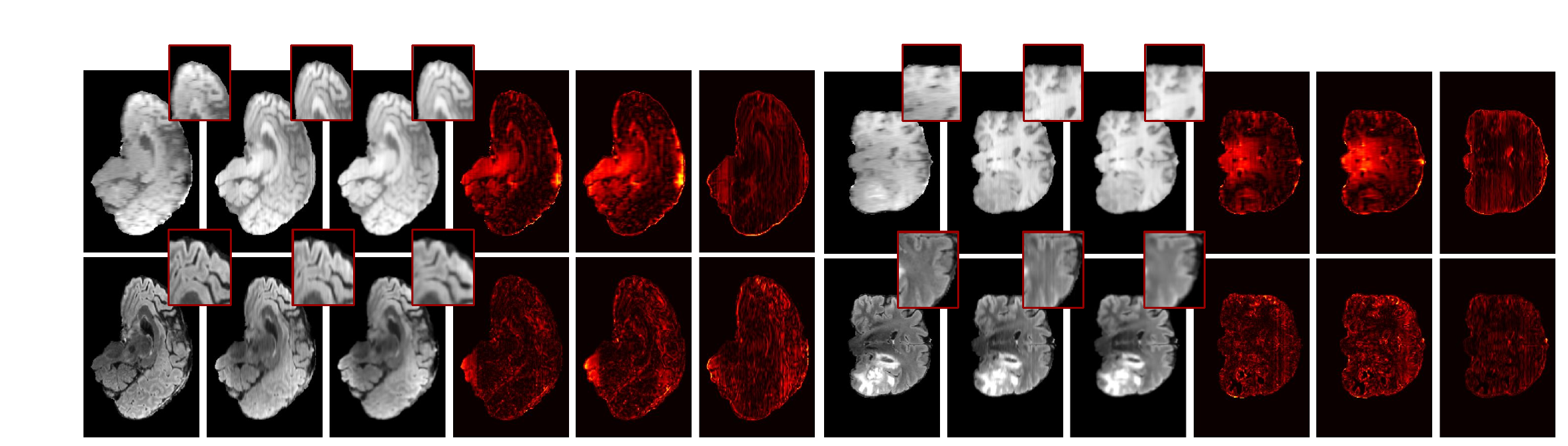}
		\put(18,28.5){\small $\text{Sagittal Plane View}$}
		\put(65,28.5){\small $\text{Coronal Plane View}$}
		\put(6,26){\tiny $\text{Original}$}
		\put(15,26){\tiny $\text{MMG}$}
		\put(23,26){\tiny $\text{CEn}$}
		\put(31,26){\tiny $\text{O-M}$}
		\put(39,26){\tiny $\text{O-C}$}
		\put(46,26){\tiny $\text{M-C}$}
		\put(53,26){\tiny $\text{Original}$}
		\put(62,26){\tiny $\text{MMG}$}
		\put(70,26){\tiny $\text{CEn}$}
		\put(78,26){\tiny $\text{O-M}$}
		\put(86,26){\tiny $\text{O-C}$}
		\put(94,26){\tiny $\text{M-C}$}
		\put(0,18){\tiny $\text{T1w}$}
		\put(0,6){\tiny $\text{Flair}$}
	\end{overpic}
	\caption{
		Qualitative comparison of coronal and sagittal plane views illustrating inter-slice inconsistency present in MMG-generated slices, which is effectively resolved in MMG$+$CEn-generated slices. Specifically, the "M–C" map highlights the removal of vertical blurry artifacts caused by inter-slice inconsistency.
	}
	\label{fig:qualitative_2}
\end{figure}

These changes may arise due to the CEn module training to enhance 3D coherence across slices, addressing potential misalignments or artifacts caused by slice-wise (2D) generation in the MMG model (see Figure \ref{fig:qualitative_2}). 
This demonstrates that while MMG excels in per-slice synthesis quality, MMG$+$CEn prioritizes spatial coherence across depth, making it more suitable for applications where 3D structural integrity is essential.
\vspace{-0.3cm}

\section{Conclusion}
\vspace{-0.3cm}
In this paper, we present SLaM-DiMM, a missing brain MRI modality generation framework comprising  MMG and CEn modules. MMG built on a diffusion-based architecture, helps learn shared latent representations across modalities, resulting in flexible generation of a desired target MRI modality. The CEn network improves inter-slice structural consistency in the generated 3D volumes, crucial for downstream 3D segmentation tasks. 
Together, these components enable robust and anatomically plausible modality synthesis, offering a promising solution for medical imaging scenarios with incomplete multimodal data.

        \section{Acknowledgment}
    We gratefully acknowledge Technocraft Centre of Applied Artificial Intelligence (TCAAI), IIT Bombay, for their support through generous funding.

\bibliographystyle{splncs04}
\bibliography{ref}

\end{document}